\documentclass{appolb} 
\usepackage{graphicx} 
\usepackage{booktabs}
\begin{document}
% \eqsec  % uncomment this line to get equations numbered by (sec.num)
\title{Proton radiography to improve proton radiotherapy: Simulation study at different proton beam energies %
\thanks{Presented at Jagiellonian Symposium on Fundamental and Applied Subatomic Physics, 7-12 June, 2015, Krak\'ow, Poland}%
% you can use '\\' to break lines
}
\author{A.K. Biegun$^{1,}$
\thanks{Corresponding author: {\it{a.k.biegun@rug.nl}}}
\address{$^{1}$KVI-Center for Advanced Radiation Technology, University of Groningen, The Netherlands}\\
{Jun Takatsu$^{2}$, M-J. van Goethem$^{3}$, E.R. van der Graaf$^{1}$, M. van Beuzekom$^{4}$, J. Visser$^{4}$, S. Brandenburg$^{1}$
}
\address{
$^{2}$Department of Radiation Oncology, Graduate School of Medicine, Osaka University, Japan\\
$^{3}$Department of Radiation Oncology, University Medical Center Groningen, University of Groningen, The Netherlands\\
$^{4}$National Institute for Subatomic Physics (Nikhef), Amsterdam, The Netherlands}
}
\maketitle
\vspace{-0.6cm}
\begin{abstract}
To improve the quality of cancer treatment with protons, a translation of X-ray Computed Tomography (CT) images into a map of the proton stopping powers 
needs to be more accurate. Proton stopping powers determined from CT images have systematic uncertainties in the calculated proton range in a patient of typically 3-4\% and even up to 10\% in region containing bone~\cite{USchneider1995,USchneider1996,WSchneider2000,GCirrone2007,HPaganetti2012,TPlautz2014,GLandry2013,JSchuemann2014}. As a consequence, part of a tumor may receive no dose, or a very high dose can be delivered in healthy ti\-ssues and organs at risks~(e.g. brain stem)~\cite{ACKnopf2013}.
A transmission radiograph of high-energy protons measuring proton stopping powers directly will allow to reduce these uncertainties, and thus improve the quality of treatment. 

The best way to obtain a sufficiently accurate radiograph is by tracking individual protons traversing the phantom (patient)~\cite{GCirrone2007,TPlautz2014,VSipala2013}. In our simulations we have used an ideal position sensitive detectors measuring a single proton before and after 
a phantom, while the residual energy of a proton was detected by a BaF$_{2}$ crystal. To obtain transmission radiographs, diffe\-rent phantom materials have been irradiated with a 3x3~cm$^{2}$ scattered proton beam, with various beam energies.
The simulations were done using the Geant4 simulation package~\cite{SAgostinelli2003}. 

In this study we focus on the simulations of the energy loss radiographs for various proton beam energies that are clinically available in proton radiotherapy. 
\end{abstract}
\PACS{42.30.Va, 87.56.-v, 87.55.Gh, 87.55.K-, 87.55.D-} 
%% https://www.aip.org/publishing/pacs/pacs-alphabetical-index#I
%Image forming, 42.30.Va 
%Radiation therapy: equipment for, 87.56.-v
%Computer modeling and simulation, 07.05.Tp: in radiation therapy, 87.55.Gh, 87.55.K-
%Treatment planning, 87.55.D-
%Radiation treatment: in medical physics, 87.55.-x
%Image reconstruction in medical imaging, 87.57.nf
\section{Introduction}
Proton radiography is one of the novel imaging modalities that has a big potential to be used in proton radiotherapy as a tool for a patient positioning and as an alternative imaging tool in proton treatment. It delivers direct information about proton stopping powers of different materials in an object through which the proton beam has passed. The image quality of a proton radiograph is reduced by the multiple Coulomb scattering and energy loss processes of protons in matter. In our study, 
we applied a cut on the proton scattering angle that optimizes the quality of the reconstructed energy loss radiographs in terms of contrast and statistical accuracy.

\section{Proton radiography setup}
\begin{figure}[htb]
	\centerline{
	\includegraphics[width=7.7cm]{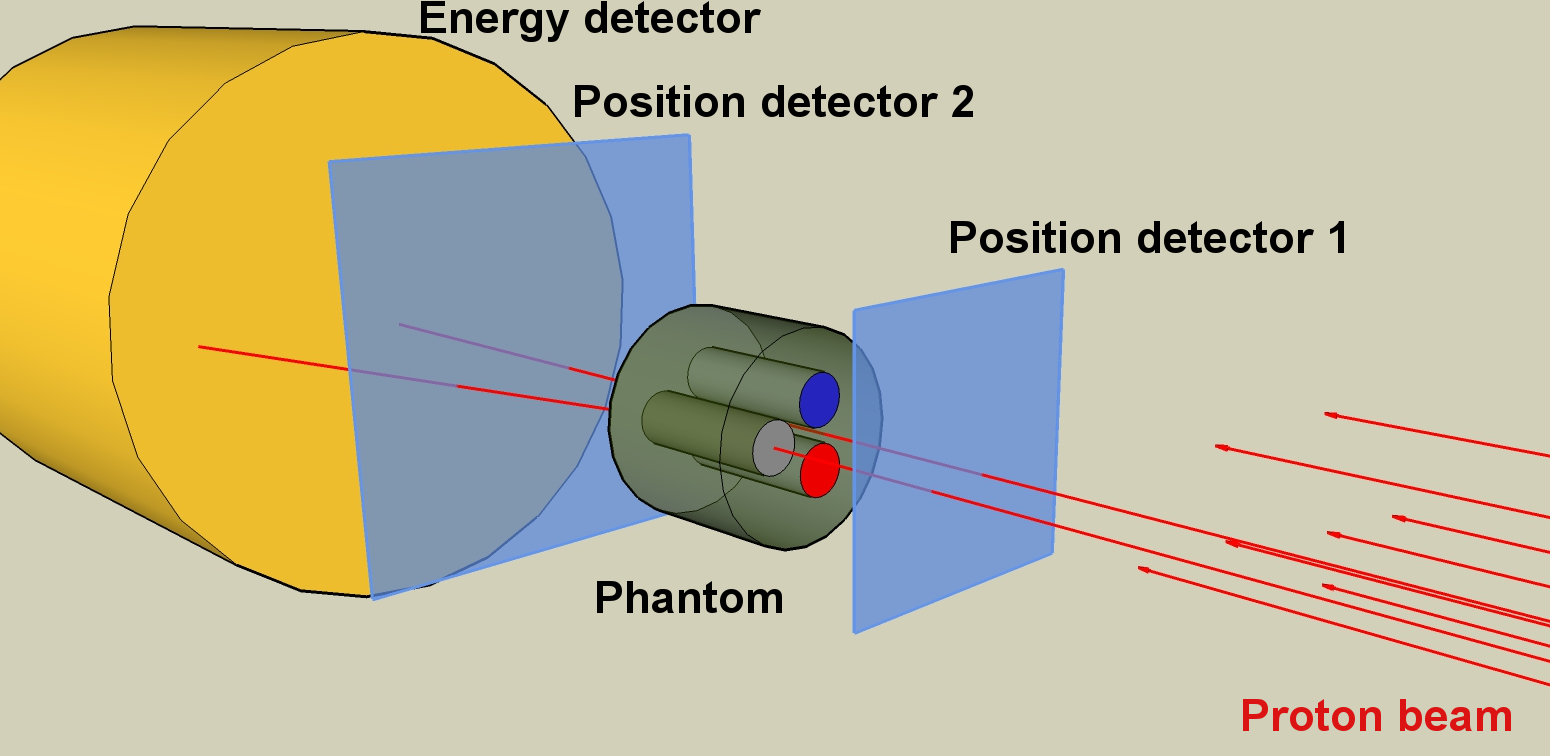}} 
	\caption{A proton radiography setup used in the Geant4 simulations. Two ideal position sensitive detectors (blue squares) and an energy detector (yellow cylinder) are shown. A scattered proton beam and a phantom containing three inserts are also presented.}
\label{G4Setup}
\end{figure}

The setup that was used to simulate the energy loss and scattering angle radiographs is presented in fig.~\ref{G4Setup}. Two ideal (100\% efficiency) position sensitive detectors with a size of 3x3~cm$^{2}$ and 10x10~cm$^{2}$ placed before and after the phantom, respectively, measured position of an individual proton. A BaF$_{2}$ energy detector with the diameter of 15~cm and a length of 15~cm was placed after the second position detector to measure the residual energy of a proton. A phantom with a size of 2.5~cm diameter and a length of 2.5~cm was located between position detectors. It was made of CT solid water 
(Gammex 457,~$\rho=1.015$~g/cm$^{3}$) and filled with PMMA ($\rho=1.19$~g/cm$^{3}$), and tissue-like materials: adipose 
(Gammex 453,~$\rho=0.92$~g/cm$^{3}$) and cortical bone (Gammex 450,~$\rho=1.82$~g/cm$^{3}$)~\cite{Gammex}. A scattered proton beam with a size of 3x3~cm$^2$ and different proton beam energies, $E_{p}=70$~MeV up to $E_{p}=230$~MeV (with a step of 20~MeV) were used to irradiate the~phantom.
\vspace{-0.4cm}    
\vspace{-0.4cm}
\section{Various proton beam energies and proton scattering angle cut}
To see the effect of the proton beam energy, $E_{p}$, on the energy radiograph of the phantom, Geant4 simulations with various proton beam energies were performed.  Proton beam energies were selected in the range available in proton radiotherapy (i.e. from $E_{p}=70$~MeV up to $E_{p}=230$~MeV). In this paper we show results for four of the selected energies: $E_{p}=90, 150, 190$ and 230~MeV. Different maximum scattering angles of the proton, such as: 17.4, 8.7, 5.2 and 1.7~mrad were applied to improve the image quality. The results for a proton beam energy of 150~MeV showed that the best trade-off between the image quality and efficiency was obtained for a proton maximum scattering angle of 5.2~mrad. For this cut nearly 50\% of the protons were used, while the image quality is almost not affected~\cite{JTakatsu2015}. For the maximum proton scattering angle of 1.7~mrad the image quality was the best, but a very high percentage of rejected protons (90.7\%) made the cut highly inefficient. 
This trend is also observed for other proton beam energies, shown in table~\ref{tab:table1}. For increasing proton beam energy and a selected angular cut the number of rejected protons decreases, and thus more protons are considered to build a proton radiograph. Number of rejected protons at angular cut of 1.7~mrad is very high, up to 82\% at $E_{p}=230$~MeV (table~\ref{tab:table1}), thus the cut remains inefficient. Therefore, in this paper the energy loss radiographs for various proton beam energies, $E_{p}$, are shown for a proton scattering angle cut of~5.2~mrad. 
\begin{table}[h!]
  \begin{center}
    \caption{Number of rejected events for maximum proton scattering angles.}
    \label{tab:table1}
        \vspace*{0.25cm}
    \begin{tabular}{ccc}
	\toprule
    	E$_{p}$ (MeV) & \hspace*{1cm}Maximum proton scattering angle \\
                   		& $<$1.7~(mrad)  & $<$5.2~(mrad) \\
       \midrule
          90       & 97.6~\%  &  82.2~\% \\
        150       & 90.7~\%  &  57.7~\% \\
        190       & 86.3~\%  &  50.3~\% \\
        230       & 81.8~\%  &  44.9~\% \\    
     \bottomrule
    \end{tabular}
  \end{center}
\end{table}

\subsection{Energy radiographs for $E_{p}=$ 90, 150, 190 and 230~MeV}
\begin{figure}[htb]
	\centerline{%
	\includegraphics[width=6.3cm]{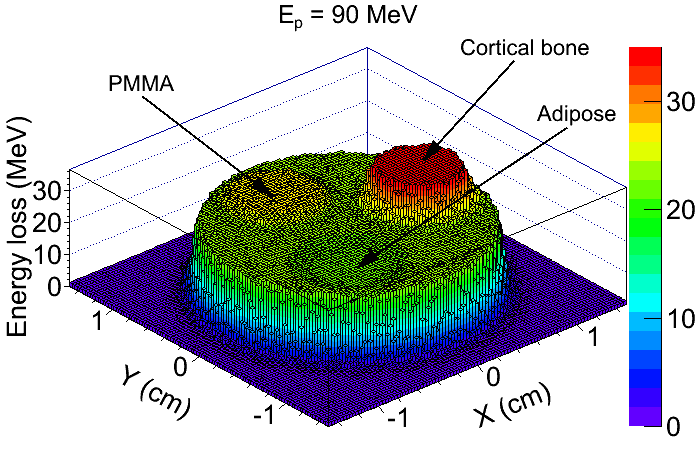}
	\includegraphics[width=6.3cm]{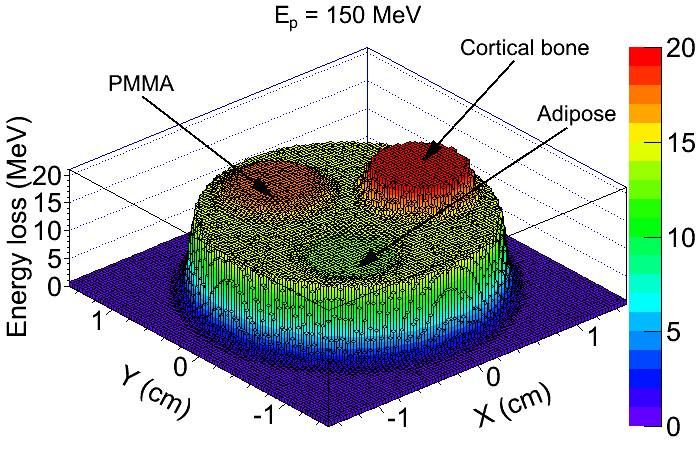}}
	\centerline{
	\includegraphics[width=6.3cm]{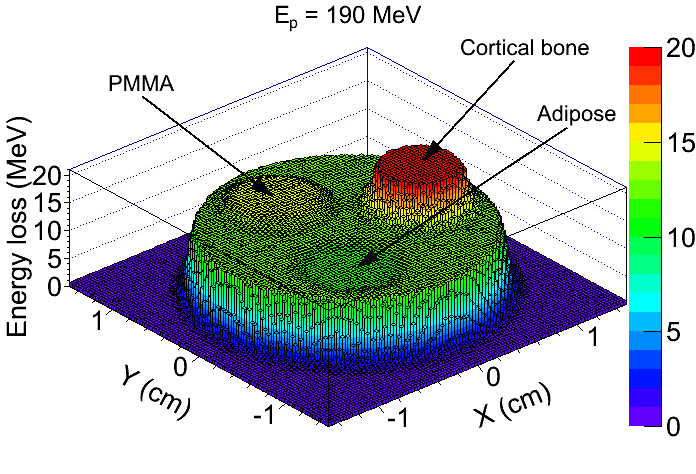}
	\includegraphics[width=6.3cm]{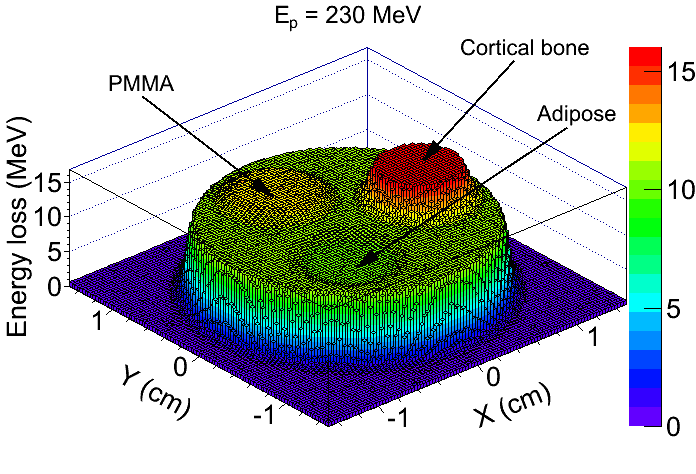}}
\caption{Proton radiographs at four proton beam energies of $E_{p}=90$~MeV (top-left), $E_{p}=150$~MeV (top-right), $E_{p}=190$~MeV (bottom-left) and $E_{p}=230$~MeV (bottom-right).  A selection of protons with the maximum scattering angle of 5.2~mrad was applied. The color scale is adjusted for better visibility of the images.}
\label{4EnRadwCut52mrad}
\end{figure}Energy radiographs at four selected proton beam energies (lower, middle and the highest available in clinics), with the angular cut of 5.2~mrad, are depicted in fig.~\ref{4EnRadwCut52mrad}. 
At all four proton beam energies and the applied angular cut, the sharp edges between materials are visible. To determine the sharpness of the boundaries between ma\-te\-rials in the phantom, projections through the phantom (in x and y directions) were evaluated.

\subsection{Projections for different proton beam energies}

\begin{figure}[htb]
	\centerline{%
	\includegraphics[width=6.35cm]{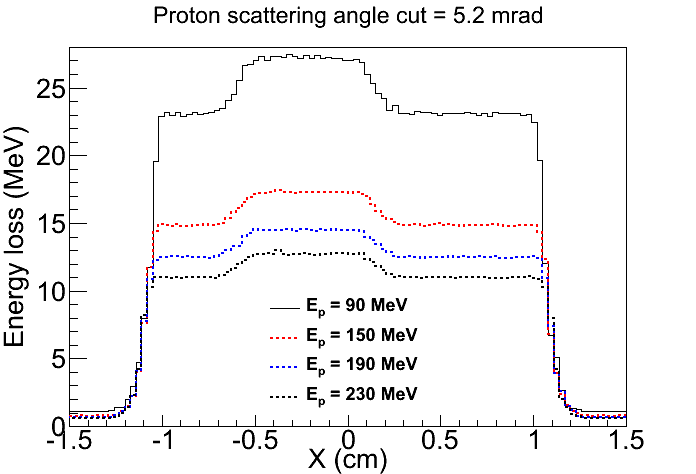}
	\includegraphics[width=6.35cm]{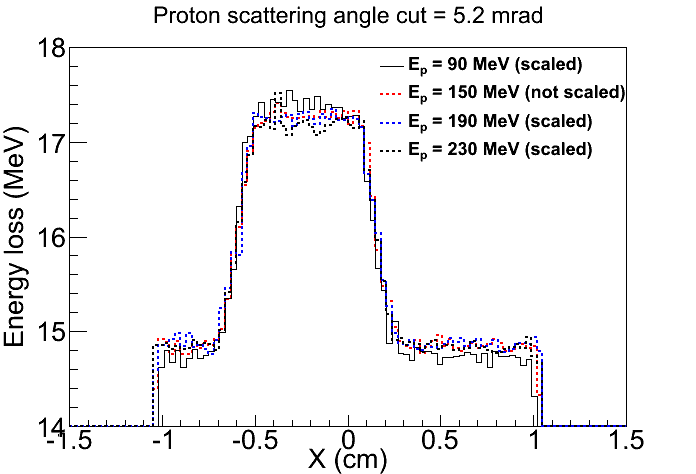}}
\caption{(Left) Projections in x direction for $y=0.5$~cm at four proton beam energies: $E_{p}=90$~MeV (black-solid line), 150~MeV (red-dashed line), 190~MeV (blue-dashed line) and 230~MeV (black-dashed line). (Right) Scaled projections from fig.~\ref{ProjXY4ProtonBeamEnergiesCut52mrad} (left) to proton beam energy of $E_{p}=150$~MeV, and zoomed between 14~MeV and 18~MeV of the energy loss.}
\label{ProjXY4ProtonBeamEnergiesCut52mrad}
\end{figure}

Projections in x-direction at $y=0.5$~cm at proton beam energies~$E_{p}=90, 150, 190$ and 230~MeV are shown in fig.~\ref{ProjXY4ProtonBeamEnergiesCut52mrad} (left). The projections were done for a single bin (bin width: 0.3~mm) with CT solid water and PMMA of proton radiographs in fig.~\ref{4EnRadwCut52mrad}.
In all projections (also in the y direction through CT solid water and cortical bone, not shown in this paper), the sharpness of the edges between materials for presented energies are comparable, as can be particularly seen in the scaled histograms in fig.~\ref{ProjXY4ProtonBeamEnergiesCut52mrad} (right). After scaling, no differences in shapes and fall-offs between materials at the four demonstrated proton beam energies are noticeable. Therefore, the angular cut of 5.2~mrad can be applied for determining edges between materials independently of the proton beam energy used for a phantom irradiation.  
\begin{figure}[htb]
	\centerline{%
	\includegraphics[width=6.3cm,height=4.8cm]{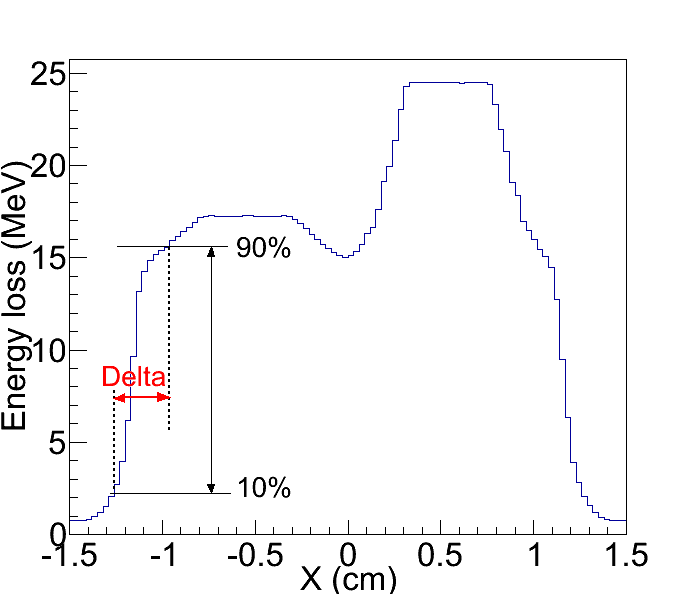}
	\includegraphics[width=6.3cm]{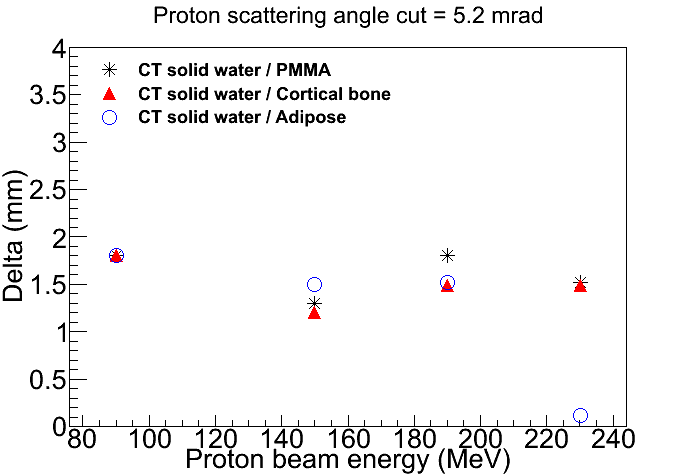}} 
	\caption{(Left) Definition of the {\it{Delta}} parameter. (Right) {\it{Delta}} parameter for different proton beam energies with the angular cut of 5.2~mrad.}
\label{Delta4ProtonBeamEnergiesCut5.2mrad}
\end{figure}

The sharpness of the boundaries between materials characterized by, so-called, {\it{Delta}} pa\-ra\-meter are calculated as a difference in position taken at 90\% and 10\% of the slope between phantom materials, such as CT solid water and either PMMA, cortical bone or adipose (fig.~\ref{Delta4ProtonBeamEnergiesCut5.2mrad}, left). For presented proton beam energies: $E_{p}=90, 150, 190$ and 230~MeV the {\it{Delta}} parameter is comparable and lower than 1.8~mm (fig.~\ref{Delta4ProtonBeamEnergiesCut5.2mrad}, right).

\section{Summary}
In this paper we analyze proton energy loss radiographs for various proton beam energies that are available in proton radiotherapy. The best energy loss radiograph with sufficient number of accepted protons at various proton beam energies was obtained for a proton scattering angle cut of 5.2~mrad. Therefore, this angular cut was applied to obtain energy loss radiographs at proton beam energies of $E_{p}=90, 150, 190$ and 230~MeV. After scaling the images, it can be seen that the edges between materials in the phantom are equally sharp for different proton beam energies (fig.~\ref{ProjXY4ProtonBeamEnergiesCut52mrad}, right), making the cut very efficient. 

Further study with a more complex phantom containing more tissue-equivalent materials and more materials inserted on the beam path, which simulates more realistic patient geometry, is being performed. 

\section*{Acknowledgement}
We would like to thank the staff of the Millipede cluster of the University of Groningen in The Netherlands, where the Monte Carlo simulations were performed (www.rug.nl/cit/hpcv/faciliteiten/HPCCluster). This work was partially supported by the Japan Society for Promotion of Science Core-to-Core Program (number 23003).

\end{document}